\documentclass[reprint,noshowpacs,noshowkeys,prd,balancelastpage,onecolumn,nofootinbib]{revtex4}
\usepackage{amsfonts}
\usepackage{mdframed}
\usepackage{amssymb}
\usepackage{footnote}
\usepackage{amsmath}
\usepackage{graphicx}
\usepackage{float}
\usepackage[font={footnotesize,it}]{caption}
\usepackage[utf8]{inputenc}
\usepackage{natbib}
\usepackage{tikz}
\usepackage{tikz-3dplot}
\usepackage[colorlinks=true,
            linkcolor=blue,
            urlcolor=blue,
            citecolor=red]{hyperref}

\begin{document}

\title{Comment on "Reissner-Nordstr\"{o}m Black Holes in the Inverse
Electrodynamics Model".}
\author{S. Habib Mazharimousavi}
\email{habib.mazhari@emu.edu.tr}
\affiliation{Department of Physics, Faculty of Arts and Sciences, Eastern Mediterranean
University, Famagusta, North Cyprus via Mersin 10, Turkey}
\date{\today }

\begin{abstract}
Recently, the inverse electrodynamics model (IEM) was introduced and applied
to fined Reissner-Nordstr\"{o}m black holes in the context of the general
relativity coupled minimally with the nonlinear electrodynamics \cite{1}.
The solution consists of both electric and magnetic fields as of the dyonic
solutions. Here in this note, we show that the IEM model belongs to a more
general class of the nonlinear electrodynamics with $T=T_{\mu }^{\mu }=0$.
Here $T_{\mu }^{\nu }$ is the energy momentum tensor of the nonlinear
electrodynamic Lagrangian. Naturally, such a dyonic RN black hole solution
is the solution for this general class.
\end{abstract}

\maketitle

In their novel work \cite{1}, Cembranos, Cruz-Dombrizb and Jarillo have
introduced the so called inverse electrodynamics model (IEM) an alternative
to the linear Maxwell theory. In their model the IEM Lagrangian density has
been proposed to be%
\begin{equation}
\mathcal{L}\left( X,Y\right) =\frac{1}{4\pi }X\left[ 1-\eta \left( \frac{Y}{X%
}\right) ^{2}\right]  \label{Lag}
\end{equation}%
in which $X$ and $Y$ are the Maxwell's invariants expressed as%
\begin{equation}
X=\frac{1}{4}F_{\mu \nu }F^{\mu \nu }
\end{equation}%
and%
\begin{equation}
Y=\frac{1}{4}F_{\mu \nu }F^{\ast \mu \nu }
\end{equation}%
with $F_{\mu \nu }=\partial _{\mu }A_{\nu }-\partial _{\nu }A_{\mu }$ the
electromagnetic field and $F_{\mu \nu }^{\ast }$ its dual. To find the
Reissner-Nordstr\"{o}m \cite{RN} black holes in the gravity coupled with the
IEM nonlinear electrodynamics, the action is chosen to be%
\begin{equation}
I=\int \sqrt{-g}\left[ \frac{\mathcal{R}}{2\kappa }+\mathcal{L}\left(
X,Y\right) \right] d^{4}x
\end{equation}%
in which $\kappa =8\pi G.$ Variation of the action with respect to the
metric tensor yields the Einstein's field equations given by ($G=c=1$)%
\begin{equation}
G_{\mu }^{\nu }=8\pi T_{\mu }^{\nu }
\end{equation}%
in which 
\begin{equation}
T_{\mu }^{\nu }=\left( \mathcal{L}-Y\mathcal{L}_{Y}\right) \delta _{\mu
}^{\nu }-\mathcal{L}_{X}F_{\mu \lambda }F^{\nu \lambda }
\end{equation}%
and a sub-index stands for the partial derivative. To find a general class
of the Lagrangian containing IEM, in the sequel, we work with a generic
Lagrangian density i.e., $\mathcal{L}\left( X,Y\right) $. Variation of the
action with respect to the electromagnetic four-potential leads to the
non-linear Maxwell equations given by%
\begin{equation}
d\left( \mathcal{L}_{X}\mathbf{F}^{\ast }+\mathcal{L}_{Y}\mathbf{F}\right) =0
\end{equation}%
in which $\mathbf{F}=\frac{1}{2}F_{\mu \nu }dx^{\mu }\wedge dx^{\nu }$ and $%
\mathbf{F}^{\ast }=\frac{1}{2}F_{\mu \nu }^{\ast }dx^{\mu }\wedge dx^{\nu }$
are the electromagnetic field two-form and its dual. Following \cite{1} we
set%
\begin{equation}
\mathbf{F}=E\left( r\right) dt\wedge dr-B\left( r\right) r^{2}\sin \theta
d\theta \wedge d\phi
\end{equation}%
with its dual field given by%
\begin{equation}
\mathbf{F}^{\ast }=-B\left( r\right) dt\wedge dr-E\left( r\right) r^{2}\sin
\theta d\theta \wedge d\phi .
\end{equation}%
The spherically symmetric line-element is considered to be%
\begin{equation}
ds^{2}=-\lambda \left( r\right) dt^{2}+\frac{1}{\lambda \left( r\right) }%
dr^{2}+r^{2}d\Omega ^{2}
\end{equation}%
upon which the Maxwell invariants are obtained to be%
\begin{equation}
X=\frac{1}{2}\left( B^{2}-E^{2}\right)
\end{equation}%
and%
\begin{equation}
Y=EB.
\end{equation}%
The Bianchi identities i.e., 
\begin{equation}
d\mathbf{F=0}
\end{equation}%
imply 
\begin{equation}
B(r)=\frac{P}{r^{2}}  \label{Magnetic}
\end{equation}%
in which $P$ is an integration constant. Following the magnetic field (\ref%
{Magnetic}), the Maxwell nonlinear equation becomes%
\begin{equation}
d\left( \mathcal{L}_{X}E\left( r\right) r^{2}+\mathcal{L}_{Y}P\right) =0.
\label{ME}
\end{equation}%
Next, we calculate the trace of the energy-momentum tensor which is given by%
\begin{equation}
T=T_{\mu }^{\mu }=\frac{1}{\pi }\left( \mathcal{L}-Y\mathcal{L}_{Y}-X%
\mathcal{L}_{X}\right) .
\end{equation}%
and then set it to zero to obtain%
\begin{equation}
\mathcal{L}-Y\mathcal{L}_{Y}-X\mathcal{L}_{X}=0  \label{PDE}
\end{equation}%
which is a nonlinear partial differential equation with respect to $X$ and $%
Y $. The most general solution to (\ref{PDE}) is derived to be%
\begin{equation}
\mathcal{L}\left( X,Y\right) =-\frac{1}{4\pi }XL\left( Z\right)
\label{Solution}
\end{equation}%
in which $Z=\frac{Y}{X}$ and $L\left( Z\right) $ is an arbitrary function of 
$Z.$ With the general solution (\ref{Solution}) one easily finds that with $%
L\left( Z\right) =1-\eta Z^{2}$ the solution coincides with the IEM given in
Eq. (\ref{Lag}). Let's continue with the general expression for the
Lagrangian density upon which the Maxwell equations(\ref{ME}) implies%
\begin{equation}
\left( L-\frac{Y}{X}L^{\prime }\right) E\left( r\right) r^{2}+LP=C
\end{equation}%
in which $C$ is an integration constant and a prime stands for the total
derivative with respect to $Z$. Explicitly one obtains $Z=\frac{2Er^{2}/P}{%
1-E^{2}r^{4}/P^{2}}$ \ which reveals that the latter equation is only a
function of $E\left( r\right) r^{2}.$ This in turn results in%
\begin{equation}
E\left( r\right) =\frac{Q}{r^{2}}
\end{equation}%
in which $Q$ is a constant satisfying the constraint%
\begin{equation}
\left( L\left( Z\right) -ZL^{\prime }\left( Z\right) \right) Q+L\left(
Z\right) P=C
\end{equation}%
with $Z=$ $\frac{2QP}{P^{2}-Q^{2}}$. We note that, having $Z$ to be $r-$%
independent, both $L$ and $L^{\prime }$ become $r-$independent and
consequently the energy-momentum tensor is found to be%
\begin{equation}
T_{\mu }^{\nu }=\frac{\xi }{8\pi r^{4}}diag\left[ -1,-1,1,1\right]
\label{EMT}
\end{equation}%
where 
\begin{equation}
\xi =\left( Q^{2}+P^{2}\right) \left( L\left( Z\right) -ZL^{\prime }\left(
Z\right) \right) .
\end{equation}%
Eq. (\ref{EMT}) suggests that the energy-momentum tensor of the general
class of the Lagrangian (\ref{Solution}) is conformal to the Maxwell linear
theory such that 
\begin{equation}
\left( T_{\mu }^{\nu }\right) _{\text{(\ref{EMT})}}=\Omega \left( T_{\mu
}^{\nu }\right) _{\text{Maxwell}}
\end{equation}%
in which $\Omega =\left( L\left( Z\right) -ZL^{\prime }\left( Z\right)
\right) $ is the conformal factor.

The $tt$-component of the Einstein's field equation becomes%
\begin{equation}
\frac{r\lambda ^{\prime }-1+\lambda }{r^{2}}=-\frac{\xi }{r^{4}}
\end{equation}%
which admits the Reissner-Nordstr\"{o}m black hole solution, i.e., 
\begin{equation}
\lambda \left( r\right) =1-\frac{2M}{r}+\frac{\xi }{r^{2}}.  \label{Metric}
\end{equation}%
We note that, with $P=0$ one simply finds $Z=0$ and consequently $\xi
=Q^{2}L\left( 0\right) ,$ which implies that, $L\left( 0\right) =1.$ This
fact can be seen clearly in the case of the IEM, where $L=1-\eta Z^{2}$. To
complete this study we investigate the least energy conditions i.e., the
weak energy conditions (WEC), which has to be satisfied by the
energy-momentum tensor. For a perfect fluid type energy-momentum tensor with 
$T_{\mu }^{\nu }=diag\left[ -\rho ,p,p,p\right] $, WEC corresponds to $\rho
\geq 0,$ and $\rho +p\geq 0.$ These conditions simplifies to $\xi \geq 0$ or
consequently $L\left( Z\right) -ZL^{\prime }\left( Z\right) \geq 0.$
Furthermore, with $\xi \geq 0$ one can show that the Strong Energy Condition
(SEC) and the Dominant Energy Condition (DEC) are also satisfied \cite{WEC}.
In IEM the quantity $\xi =\left( Q^{2}+P^{2}\right) \left( 1+\eta
Z^{2}\right) $ which is positive definite if and only if $\eta \geq 0.$
Introducing $q^{2}=\xi \geq 0$ the metric (\ref{Metric}) becomes%
\begin{equation*}
\lambda \left( r\right) =1-\frac{2M}{r}+\frac{q^{2}}{r^{2}}
\end{equation*}%
which possesses exactly the same properties as of the Reissner-Nordstr\"{o}m
black hole in the Einstein-Maxwell theory.

In conclusion, we reconsidered the IEM model of nonlinear electrodynamics
introduced in \cite{1}. It has been shown that, IEM model belongs to a
general class of nonlinear electrodynamics admitting the trace of the
corresponding energy-momentum tensor zero. The general expression of the
model has been explicitly obtained as of Eq. (\ref{Solution}). It is worth
to mention that, in \cite{2} and \cite{3}, similar formalism has been
developed through introducing the conformally invariant Maxwell source in
the context of different forms for the Maxwell's invariants. However, the
idea of the conformal nonlinear electrodynamics and the traceless
energy-momentum tensor have been known earlier in the works of Cataldo et
al. \cite{4} and Hassaine and Martinez \cite{5}. Finally, we would like to
mention the two very recent works which investigate the NED theory
preserving both conformal invariance and $SO(2)$ duality-rotation invariance 
\cite{SO2}.

\end{document}